\documentclass[conference]{IEEEtran}
\IEEEoverridecommandlockouts
\usepackage{url}
\usepackage{diagbox}
\usepackage{cite}
\usepackage{amsmath,amssymb,amsfonts}
\usepackage{algorithmic}
\usepackage{graphicx}
\usepackage{textcomp}
\usepackage{xcolor}
\usepackage{soul}
\usepackage[a4paper, total={184mm,239mm}]{geometry}
\def\BibTeX{{\rm B\kern-.05em{\sc i\kern-.025em b}\kern-.08em
    T\kern-.1667em\lower.7ex\hbox{E}\kern-.125emX}}

\long\def\comment#1{}
\newcommand{\RVPKR}{\texttt{PKR}}
\newcommand{\WRPKRU}{\texttt{WRPKR}}
\newcommand{\RDPKRU}{\texttt{RDPKR}}
\newcommand{\numSpec}{6}
\newcommand{\numSpecSix}{4}
\newcommand{\numMibench}{7}
\newcommand{\numSpecRemaining}{4}
\newcommand{\numSpecSixRemaining}{6}
\newcommand{\Inline}{\texttt{Inline}}
\newcommand{\Func}{\texttt{Func}}
\newcommand{\MPK}{\texttt{SealPK-WR}}
\newcommand{\MPKR}{\texttt{SealPK-RD+RW}}
\newcommand{\mprotect}{\texttt{mprotect}}
\newcommand{\RVMPK}{SealPK}
\newcommand{\SealReg}{\texttt{SealReg}}
\newcommand{\PKCAM}{\texttt{PK-CAM}}
\newcommand{\sealstart}{\texttt{seal\_start}}
\newcommand{\sealend}{\texttt{seal\_end}}

\newcommand{\psMPKR}{21.00\%}
\newcommand{\psmprotect}{2875.62\%}

\newcommand{\psixMPKR}{14.81\%}
\newcommand{\psixmprotect}{1982.70\%}

\newcommand{\pmMPKR}{8.52\%}
\newcommand{\pmmprotect}{320.21\%}

\newcommand{\LUTs}{5.62\%}
\newcommand{\FFs}{2.72\%}

\newcommand{\pMPKmprotectTimes}{$88\times$}

\newcommand{\ignore}[1]{}

\usepackage{enumitem}
\usepackage{booktabs}
\usepackage{adjustbox}
\usepackage{multirow}

\usepackage{listings}
\usepackage{xcolor}

\definecolor{codegreen}{rgb}{0,0.6,0}
\definecolor{codegray}{rgb}{0.5,0.5,0.5}
\definecolor{codepurple}{rgb}{0.58,0,0.82}
\definecolor{backcolour}{rgb}{0.95,0.95,0.92}

\lstdefinestyle{mystyle}{
    backgroundcolor=\color{backcolour},   
    commentstyle=\color{codegreen},
    keywordstyle=\color{magenta},
    numberstyle=\tiny\color{codegray},
    stringstyle=\color{codepurple},
    basicstyle=\ttfamily\footnotesize,
    breakatwhitespace=false,         
    breaklines=true,                 
    captionpos=b,                    
    keepspaces=true,                 
    numbers=left,                    
    numbersep=5pt,                  
    showspaces=false,                
    showstringspaces=false,
    showtabs=false,                  
    tabsize=2
}

\ifCLASSOPTIONcompsoc
  \usepackage[nocompress]{cite}
\else
  \usepackage{cite}
\fi

\begin{document}
\title{Efficient Sealable Protection Keys for RISC-V}

\author{\IEEEauthorblockN{Leila Delshadtehrani, Sadullah Canakci, Manuel Egele, and Ajay Joshi}\\ 
  \IEEEauthorblockA{Department of Electrical and Computer Engineering, Boston University\\ 
    \{delshad, scanakci, megele, joshi\}@bu.edu}}

\maketitle

\begin{abstract}
With the continuous increase in the number of software-based attacks, there has been a growing effort towards isolating sensitive data and trusted software components from untrusted third-party components.
A hardware-assisted intra-process isolation mechanism enables software developers to partition a process into isolated components and in turn secure sensitive data from untrusted components.
However, most of the existing hardware-assisted intra-process isolation mechanisms in modern processors, such as ARM and IBM Power, rely on costly kernel operations for switching between trusted and untrusted domains.
Recently, Intel introduced a new hardware feature for intra-process memory isolation, called Memory Protection Keys (MPK), which enables a user-space process to switch the domains in an efficient way.
While the efficiency of Intel MPK enables developers to leverage it for common use cases such as Code-Pointer Integrity, the limited number of unique domains (16) prohibits its use in cases such as OpenSSL where a large number of domains are required.
Moreover, Intel MPK suffers from the protection key use-after-free vulnerability.
To address these shortcomings, in this paper, we propose an efficient intra-process isolation technique for the RISC-V open ISA, called {\RVMPK}, which supports up to 1024 unique domains.
{\RVMPK} prevents the protection key use-after-free problem by leveraging a lazy de-allocation approach.
To further strengthen \RVMPK, we devise three novel sealing features to protect the allocated domains, their associated pages, and their permissions from modifications or tampering by an attacker.
To demonstrate the feasibility of our design, we implement {\RVMPK} on a RISC-V Rocket processor, provide the OS support for it, and prototype our design on an FPGA.
We demonstrate the efficiency of {\RVMPK} by leveraging it to implement an isolated shadow stack on our FPGA prototype.
\end{abstract}
\begin{IEEEkeywords}
Intra-Process Memory Isolation, Memory Protection Keys, RISC-V, Isolated Shadow Stack
\end{IEEEkeywords}

\IEEEpeerreviewmaketitle

\section{Introduction}
\label{sec:intro}
With the ever-increasing complexity of software applications, today's software code consists of both trusted components designed in-house and untrusted components such as third-party libraries and application plugins.
The coexistence of trusted components with potentially malicious or vulnerable untrusted components in the same address space could compromise the security of the system through information leakage, denial-of-service attack, etc~\cite{vilanova2014codoms}.
While the user-space inter-process isolation protects processes from one another, the intra-process isolation of various software components has been a challenge.
Although it is feasible to invoke the \texttt{mprotect} system call from the user space to update the permission bits of specific pages, the performance overhead of \texttt{mprotect} due to the context switches between the kernel and user-space can be prohibitive (1,094 cycles on avg. on a modern processor~\cite{park2019libmpk}).

To facilitate the intra-process memory protection, in recent years, some processors such as ARM~\cite{arm2018ref} and IBM Power~\cite{ibm2017ref} have provided new features to create memory domains by assigning the same \texttt{key} to a group of memory pages.
However, these features still rely on costly kernel operations for changing domains.
More recently, Intel proposed a similar hardware feature, called Intel Memory Protection Keys (MPK)~\cite{intel2019mpk}, to efficiently support intra-process memory isolation using a user-space instruction (\texttt{WRPKRU)} to update the associated permission of a domain.
Intel MPK allows the user to create a protection domain by assigning a protection key (pkey) to a group of memory pages.
The non-privileged \texttt{WRPKRU} instruction, which updates the pkey permissions, takes about 11-260 cycles~\cite{vahldiek2019erim}, does not require a context switch, and does not lead to a TLB flush.
However, Intel MPK suffers from two major drawbacks, i.e., security and scalability.
In terms of security, Intel MPK suffers from pkey use-after-free vulnerability~\cite{park2019libmpk}.
Once a pkey gets freed, the kernel does not update the pkey bits of its associated pages. 
The same freed pkey can later on be allocated to a new domain; as a result, the old pages and the new ones will unintentionally share the same pkey.
Additionally, if an attacker tampers with a protection domain, its associated pages, or its corresponding permission, the protection keys serve no purpose.
In particular, since Intel MPK allows a user-space code to modify the pkey permissions, a malicious component might contain \texttt{WRPKRU} instructions or inject those instructions at run-time to update the permission bits of a domain and attain access to a protected domain.
In terms of scalability, Intel MPK provides only 16 pkeys.
However, some real-world use cases such as Persistent Memory Object (PMO)~\cite{xu2020hardware} and OpenSSL~\cite{park2019libmpk} require more than 1000 pkeys.
The above-mentioned drawbacks hinder the deployment of Intel MPK for enforcing granular intra-process memory isolation.

To enable pervasive deployment of Intel MPK, recent works have focused on addressing one or more of the above-mentioned drawbacks.
To prevent the manipulation of a domain's permissions by an attacker, recent works have leveraged binary inspection and binary rewriting approaches~\cite{vahldiek2019erim, hedayati2019hodor}.
Although Control-Flow Integrity (CFI)~\cite{abadi2009control} can also be utilized to prohibit an uncontrolled execution of \texttt{WRPKRU} instruction ~\cite{park2019libmpk}, CFI enforcement mechanisms incur considerable performance overhead.
To address the limited number of pkeys, libmpk~ \cite{park2019libmpk} and Xu et al.~\cite{xu2020hardware} have proposed a software-based and a hardware-based virtualization technique, respectively.
The virtualization technique of libmpk suffers from large overheads due to expensive Page Table Entry (PTE) updates~\cite{xu2020hardware}.
While the hardware-based virtualization technique by Xu et al.~\cite{xu2020hardware} provides an efficient implementation, it is not generic and is tailored for a specific application (PMO protection).
Note that the pkey virtualization techniques can eliminate the pkey use-after-free problem. 

In this paper, we propose an efficient intra-process memory isolation capability, called {\RVMPK}, leveraging the Open RISC-V Instruction Set Architecture (ISA)~\cite{waterman2011risc}.
Similar to Intel MPK, {\RVMPK} provides a per-page protection key; however, {\RVMPK} supports up to 1024 domains (64${\times}$ more than Intel MPK) by leveraging the 10 unused bits available in the PTE of each virtual page (Sv-39).
We eliminate the pkey use-after-free problem at Operating System (OS) level by keeping track of the number of pages belonging to the same domain and a lazy de-allocation approach. 
While Intel MPK does not provide a solution to maintain the integrity of protection domains and their permissions, we propose three novel sealing features to prevent an attacker from modifying sealed domains, their corresponding sealed pages, and their permissions.
In particular, our hardware-assisted permission sealing feature enables the software developer to restrict the access to \texttt{WRPKRU} within a contiguous range of memory addresses, e.g., a trusted component.
Any attempt to execute a \texttt{WRPKRU} instruction outside of the specified range would lead to a hardware exception.
Hence, this sealing feature efficiently prevents the manipulation of a domain's permission by an attacker.
To summarize, our contributions are as follows:
\begin{itemize}
    \item We present an efficient intra-process isolation capability, called {\RVMPK}, which supports up to 1024 unique isolated domains.
    We propose an OS-level solution to avoid the pkey use-after-free issue.
    We devise three novel sealing features to protect the domains, their associated pages, and their permissions from unauthorized modifications.
    \item We implement {\RVMPK} on a RISC-V Rocket processor~\cite{asanovic2016rocket} and extend the Linux kernel to support the protection keys for the RISC-V ISA.
    We evaluate a prototype of our hardware design on an FPGA with a full Linux software stack.
    \item We demonstrate the efficiency of our design by implementing an isolated shadow stack leveraging {\RVMPK}.
    Our isolated shadow stack prototype is, on average, $\sim${\pMPKmprotectTimes} faster than an isolated implementation using mprotect across SPECint2000, SPECint2006, and MiBench benchmarks.
\end{itemize}

\section{Background}
\label{sec:background}
\subsection{Memory Protection Keys}
In recent years, a growing number of modern processors have provided a per-page protection key capability, where a group of virtual memory pages form a domain and all pages in the domain are assigned the same protection key.
Intel MPK utilizes 4 previously unused bits of the PTE to specify the pkey of each page and to divide the address space into up to 16 different protection domains.
Intel MPK stores the permission bits of all the pkeys in a single 32-bit register (per logical core),
called protection key rights register (\texttt{PKRU}).
The access permission of each pkey is specified using a 2-bit value in the \texttt{PKRU}.
Accordingly, each pkey specifies a domain as \texttt{readable/writable}, \texttt{read-only} or \texttt{non-accessible}.

Intel MPK provides two new unprivileged instructions, i.e., \texttt{WRPKRU} and \texttt{RDPKRU}, to write/read into/from \texttt{PKRU}.
A user can leverage the \texttt{WRPKRU} instruction to update the permission bits of all domains without the need for a context switch.
Hence, updating the permission bits of a domain is fast (11-260 cycles~\cite{vahldiek2019erim}); however, \texttt{PKRU} is not protected from manipulation by control-flow hijacking attacks~\cite{vahldiek2019erim,hedayati2019hodor,schrammel2020donky}.

The Linux kernel provides support for Intel MPK (since v4.6) through three new system calls, i.e., \texttt{pkey\_alloc}, \texttt{pkey\_free}, and \texttt{pkey\_mprotect}.
The kernel maintains a 16-bit allocation bitmap to keep track of the allocated keys.
A user-space thread has to allocate a new pkey using the \texttt{pkey\_alloc} system call prior to assigning the pkey to a page (group) by invoking the \texttt{pkey\_mprotect} system call.
Using the \texttt{pkey\_free} system call, the user frees an allocated pkey; however, the kernel only
updates the allocation map to indicate that the corresponding pkey is free without erasing the pkey from
the PTE of all the corresponding memory pages.
The same pkey might be assigned to another domain on future \texttt{pkey\_alloc} invocations; hence,
unintentionally the previous domain would share the same pkey as the new domain, giving rise to the pkey use-after-free problem.

\subsection{RISC-V}
\label{sec:bg-riscv}
\begin{figure}
  \centering
  \includegraphics[width=0.95\linewidth] {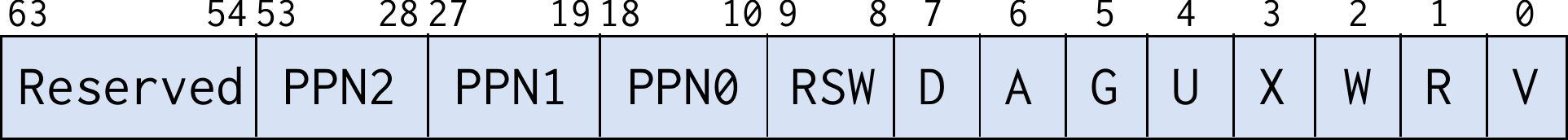}
  \caption{Sv39 PTE according to the RISC-V ISA~\cite{waterman2015risc}.}
  \label{fig:sv39}
\end{figure}

The RISC-V Instruction Set Architecture (ISA)~\cite{waterman2011risc} is an open ISA.
As part of the privileged ISA, RISC-V specifies a page-based 39-bit virtual memory, i.e., Sv39, for 64-bit systems~\cite{waterman2015risc}.
Figure~\ref{fig:sv39} shows the PTE bits of Sv39, where bits 1-3 (R, W, and X bits) are the permission bits of the page, indicating whether the page is \texttt{readable}, \texttt{writable}, and \texttt{executable}, respectively.
As shown in Figure~\ref{fig:sv39}, bits 54-63 are currently unused and reserved for future use; hence, as we will discuss in Section~\ref{sec:design}, we leverage these 10 bits to store our per-page pkey.

\section{{\RVMPK}: Design}
\label{sec:design}
In this section, we explain the baseline hardware design of {\RVMPK} and its OS support.
We discuss the design and implementation of the three novel features of {\RVMPK} in Section~\ref{sec:sealable}.
\subsection{Hardware Design}
\begin{figure}[t]
  \centering
  \includegraphics[width=0.95\linewidth] {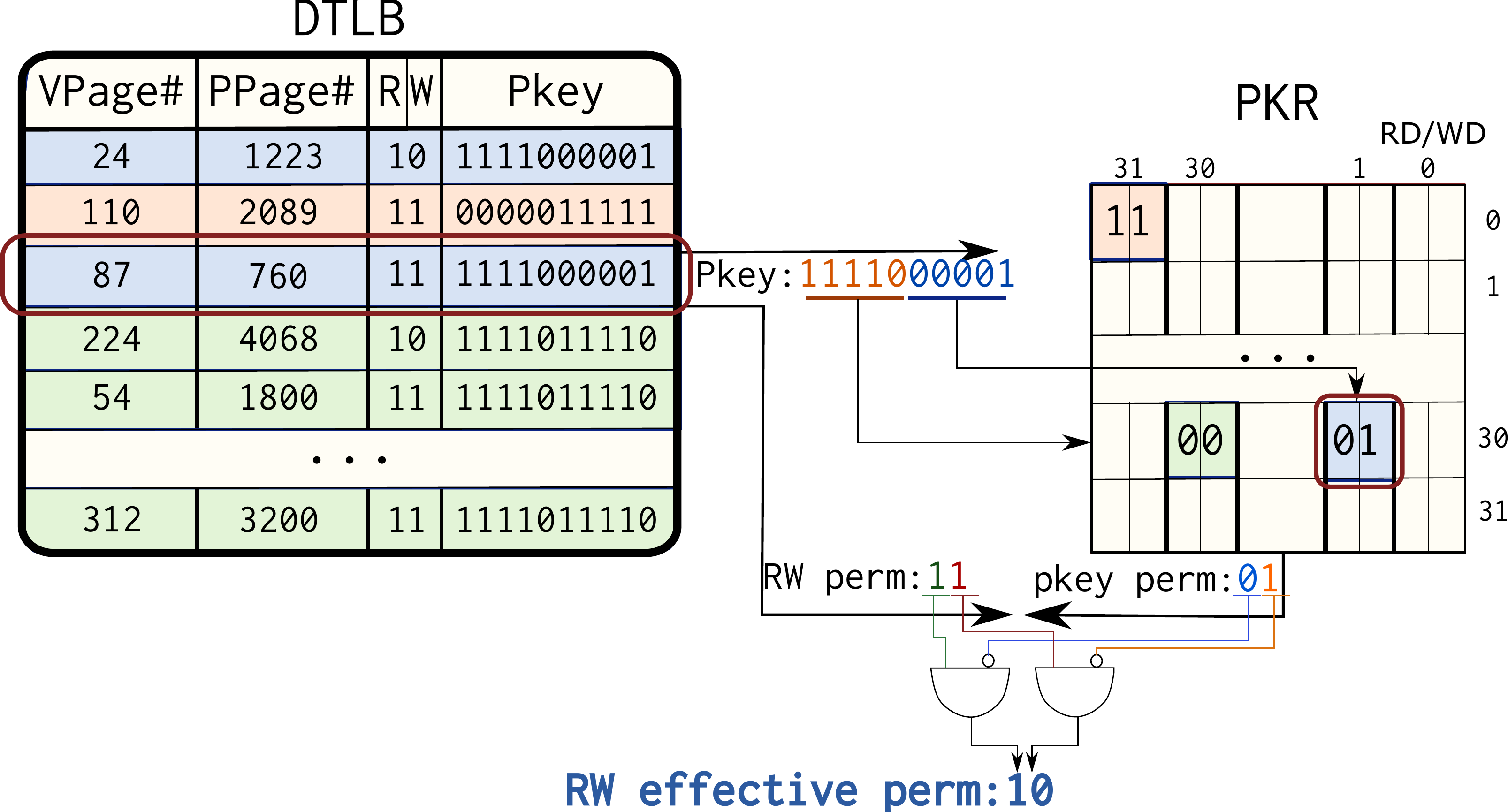}
  \caption{Modified MMU of the RISC-V Rocket core.
  Here, we color-code the TLB entries of each domain, consisting of various pages sharing the same pkey.
  For each data memory access, the effective permission bits are determined by the intersection of the PTE permissions and pkey permissions stored in {\RVPKR}.}
  \label{fig:PKRU}
\end{figure}

As mentioned before, scalability is one of the limitations of Intel MPK, as it cannot support more than 16 pkeys.
In RISC-V, we leverage the 10 unused bits (which enables up to 1024 pkeys) of the Sv39 PTE to store the pkey.\footnote{Note that the Sv48 PTE also has 10 unused bits while a 32-bit RISC-V processor uses Sv32, where there are no unused bits in PTE.
In this case, we can store the pkey information in a separate OS-managed data structure and use a TLB to cache the information at hardware level.}
Figure~\ref{fig:PKRU} demonstrates our hardware modifications to support {\RVMPK}.
We add a new entry to each line of the DTLB to store the corresponding 10-bit pkey of each virtual page.\footnote{Note that pkey checks are only applicable to data memory accesses and not an instruction fetch.
Hence, we do not modify the ITLB.
}
Hence, our {\RVMPK} design supports up to 1024 domains, which is 64$\times$ more than the 16 domains supported by Intel MPK.
We can use a virtualization-based mechanism, like libmpk~\cite{park2019libmpk}, to support more than 1024 domains.
Note that with a virtualization technique, we can create more than 1024 domains, but in reality we are still limited to 1024 concurrent physical pkeys. 
For an \texttt{unlimited} number of domains, we can store the pkey information in a separate OS-managed hierarchical structure.
We store the permission bits of the pkeys separately.
In our design, we use 2 bits, i.e., \texttt{(\texttt{Read Disable} (RD), \texttt{Write Disable} (WD))}, to specify the access permission of each protection key. 
Following the principle of the least privilege, unlike Intel MPK and previous works, our design enables a \texttt{write-only} page, which can in turn reduce the attack surface.
Such a \texttt{write-only} page is specifically useful for log entries, where one thread is responsible for writing the log and another thread processes the written log.
Note that the RISC-V ISA does not support \texttt{write-only} pages,\footnote{The PTE permissions for a \texttt{write-only} page is a feature reserved for the future use.} and our design provides this feature by leveraging pkeys regardless of the support in PTE permissions.

We support 1024 pkeys in our design; hence, unlike Intel MPK, we cannot simply use a single register to store all the pkey permission bits.
To provide fast access to these bits, we use a 2Kb on-chip SRAM-based memory to store the permission bits.
This memory, called {\RVPKR} (shown in Figure~\ref{fig:PKRU}), consists of 32 rows, where each row stores the permission bits of 32 pkeys (64 bits total).
We utilize the custom instruction extension of the RISC-V ISA~\cite{waterman2011risc} to define two new instructions, {\RDPKRU} and {\WRPKRU}, to read from and write to {\RVPKR}.\footnote{To simplify the implementation of the custom instructions, we leverage the RoCC extension of the Rocket core, which adds the support for decoding and executing custom instructions to the Rocket core's pipeline.}
The {\RDPKRU} instruction uses two registers, i.e., \texttt{rs1} and \texttt{rd}, for its operation. The input register (\texttt{rs1}) contains the pkey. At hardware level, the upper 5-bits of the pkey are used to index into {\RVPKR} and read the corresponding 64-bit row of permissions.
This 64-bit value is returned as the output and stored in \texttt{rd}.
The {\WRPKRU} instruction uses two input registers, i.e., \texttt{rs1} and \texttt{rs2}, for its operation. The first input register (\texttt{rs1}) contains the pkey, which is used to index into {\RVPKR}.
The second input register (\texttt{rs2}) contains the new value of 64-bit permissions of the corresponding row.
The hardware uses this new 64-bit value to overwrite the permission bits of the row indexed by pkey.

In our hardware design, we provide a control logic to determine the effective permission bits of each data memory access.
Consider the example shown in Figure~\ref{fig:PKRU}, where there is an incoming write request to the virtual page \#87.
In addition to reading the page's read/write permission bits stored in DTLB (\texttt{11}), the control logic reads the corresponding 2-bit permission bits of the pkey (\texttt{111100001}) stored in {\RVPKR}.
The control logic uses the upper 5 bits of the pkey to index into a specific 64-bit row of {\RVPKR} and the lower 5 bits to select the 2 permission bits (\texttt{01}).
The effective permission is the intersection of the DTLB's and pkey's permission bits.
In this example, the effective permission is \texttt{10}; hence, the write access is not allowed.
If a data access is not allowed according to the effective permission, it leads to a load/store page fault; the processor triggers an exception, and the OS handles the page fault.
\subsection{Kernel Support}
\label{sec:OS}
At the OS level, we add the support to store each page's pkey in the 10 unused bits of the PTE.
Our RISC-V kernel support is built upon the existing Linux kernel support for MPK.

\subsubsection{Lazy de-allocation}
To keep track of the allocated pkeys, we implement a 1024-bit allocation bitmap.
To efficiently address the pkey use-after-free problem of Intel MPK, we leverage a lazy de-allocation approach.
We implement a 1024-bit dirty map to indicate whether each pkey has been lazily de-allocated.
We also keep track of the number of pages currently associated with each pkey using a counter map.
If a pkey's corresponding counter is not zero, \texttt{pkey\_free} updates the permission bits of the pkey in {\RVPKR} to \texttt{(0,0)}; hence, the page-table permissions determine the effective permission of the corresponding pages.
Rather than clearing the corresponding bit of the pkey in the allocation map, \texttt{pkey\_free} sets the dirty bit and \texttt{pkey\_alloc} would not allocate a dirty pkey.
Whenever a memory page with a dirty pkey gets freed, we update the number of pages associated with the dirty pkey in the counter map, accordingly.
Once the counter becomes zero, we erase the dirty bit of the corresponding pkey; hence, it can safely be allocated afterwards.
If \texttt{pkey\_alloc} cannot find a free non-dirty pkey, it returns an allocation error to indicate no free pkey is available.

\subsubsection{Per thread OS support}
We modify the \texttt{task\_struct} in the Linux kernel to maintain the contents of {\RVPKR} for each thread during the context switches.\footnote{According to our evaluations, maintaining {\RVPKR} information during context switches incurs less than 1\% performance overhead.}
Furthermore, we modify the RISC-V page fault handler in the Linux kernel to identify a page fault caused by a pkey permission violation.
We augment the segmentation fault with the pkey information to accurately reflect the cause of the fault to the developer and assist with debugging.
\vspace{-5pt}
\section{{\RVMPK}: Sealing Features}
\label{sec:sealable}
\begin{figure}[t]
  \centering
  \includegraphics[width=0.98\linewidth] {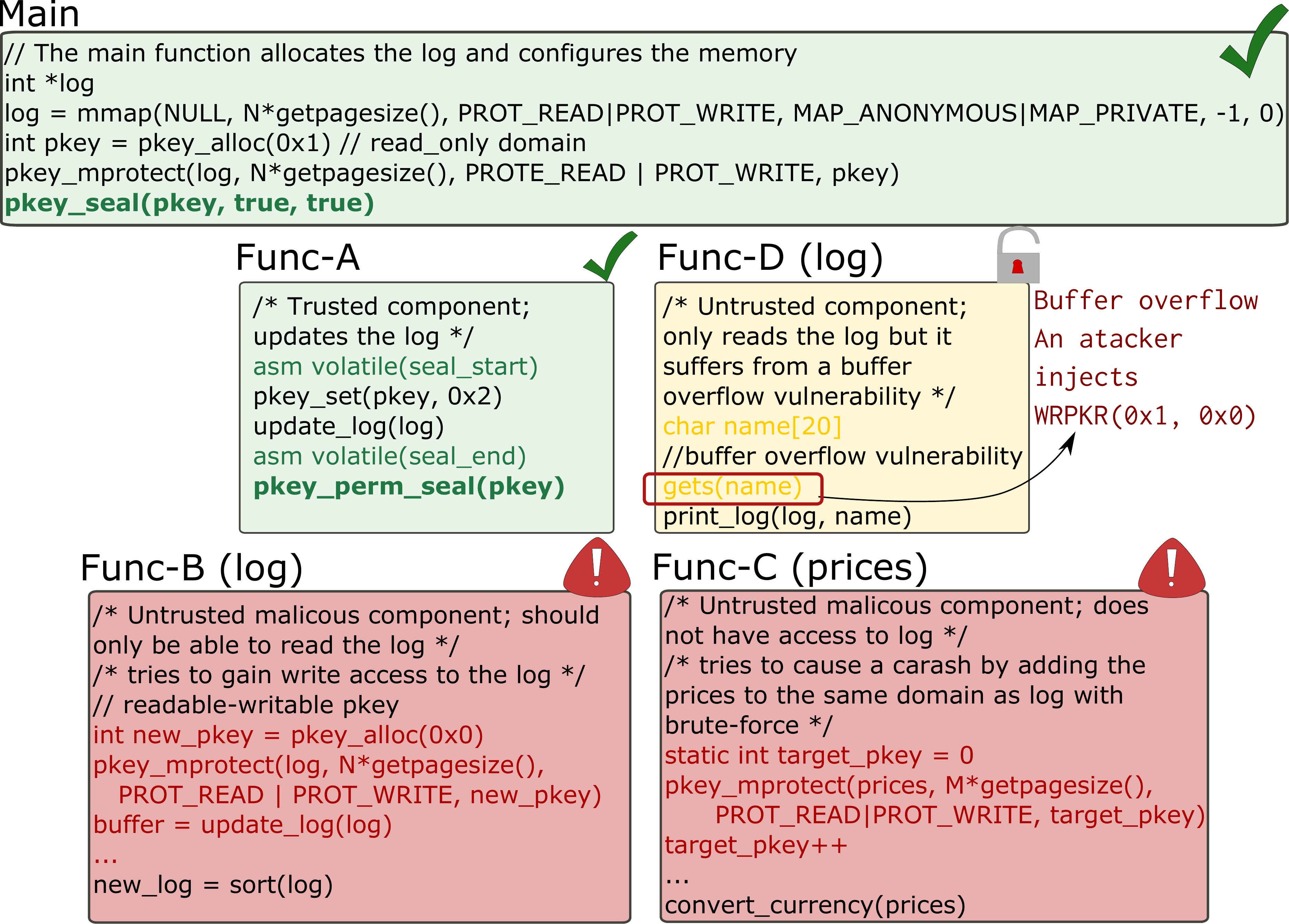}
  \caption{Example scenario for our sealable features:
   The red texts in \texttt{Func-B} and \texttt{Func-C} show an effort to attack the pkeys, the yellow texts in \texttt{Func-D} show a vulnerability that can be leveraged by an attacker to compromise the pkey permissions.
   The green texts in the \texttt{Main} and \texttt{Func-A} functions show our sealing features to protect the domain, its associated pages, and its permissions from unauthorized modifications.}
  \label{fig:log}
\end{figure}

As mentioned before, Intel  MPK  does  not  protect the allocated domains, their associated pages, and their permission bits from tampering by an attacker.
In this section, we describe three novel sealing features to protect against such tampering.
To clarify the defensive capabilities of these features, consider the example shown in Figure~\ref{fig:log}.
In this example, a software developer writes a program that handles sensitive financial records.
The \texttt{Main} function (written in-house) initially allocates the memory pages for the financial record (\texttt{log}) as \texttt{readable-writable} and assigns a protection key to these pages.
Following the principle of the least privilege, the initial value of the pkey restricts the permission to \texttt{read-only} pages.
In this example, \texttt{Func-A} updates the contents of the \texttt{log}.
We assume that this function is developed in-house and has access to the pkey.
Prior to writing the sensitive financial information into the \texttt{log}, \texttt{Func-A} modifies the domain permission of the \texttt{log} to \texttt{write-only}.
For performance reasons, the software developer leverages third-party untrusted libraries in the implementation of \texttt{Func-B}, \texttt{Func-C}, and \texttt{Func-D}.
\texttt{Funct-B} reads the \texttt{log} and returns a sorted copy of the \texttt{log}.
\texttt{Func-C} does not have access to the \texttt{log}, instead it receives a list of prices and converts them to a different currency. 
\texttt{Funct-D} reads the \texttt{log} and prints all the transactions of a specific account.
Hence, \texttt{Func-B} and \texttt{Func-D}, can only access the \texttt{log} as \textbf{read-only} memory.
For security reasons, the untrusted functions are not aware of the pkey value.
In the rest of this section, we explain how each of our sealing features protects the \texttt{log} against potential attacks originating from the untrusted components.

\textbf{Sealing the domain:} In this scenario, \texttt{Func-B} is a malicious third-party component, which receives the \texttt{log} as a \texttt{read-only} input.
\texttt{Funct-B} is supposed to read the \texttt{log} and return a sorted copy of it.
However, as shown in Figure~\ref{fig:log}, this untrusted component allocates a new \texttt{readable-writable} pkey, invokes the \texttt{mprotect} system call and assigns the new pkey to the \texttt{log}.
In this way, \texttt{Func-B} can falsify the financial records stored in the \texttt{log}.
Unfortunately, the developer who uses this untrusted function does not have access to its source-code and is unaware of its maliciousness.
In this scenario, Intel MPK is not capable of preventing this malicious modification to the \texttt{log} within the same thread.
To prevent such unauthorized modifications, we provide a domain sealing option by adding a \texttt{sealed\_domain} map to the kernel.
We modify the \texttt{pkey\_mprotect} system call to check the \texttt{sealed\_domain} map prior to modifying a domain's pkey.
Once a domain is sealed, \texttt{pkey\_mprotect} prevents any further modifications to PTE permissions as well as the pkey value, efficiently throwing such attacks.

\textbf{Sealing pages:} We assume that after the initialization step in the \texttt{Main} function, no more pages will be added to the protection domain.
Consider a scenario where \texttt{Func-C}, a malicious third-party component, aims to crash this financial application.
Crashing the application at run-time could lead to denial-of-service and financial losses.
\texttt{Func-C} does not have access to the \texttt{log}; it only receives a list of prices and converts them from one currency to another one.
This price list does not include any sensitive information; hence, \texttt{Func-A} does not assign a protection domain to it.
In this example, in each call, the malicious \texttt{Func-C} adds the pages associated with the price list to a different domain, hoping that the new domain would restrict the \texttt{read} permission.
As a result, after the price list is assigned with the same pkey as the \texttt{log}, once \texttt{Func-A} tries to read the price list the program crashes with a segmentation fault.
Intel MPK cannot prevent this issue within the same thread; similarly, our domain sealing feature is not sufficient in this scenario.
To ensure that no more pages can be added to a domain (either by mistake or by a malicious component), we provide a page sealing option by adding a \texttt{sealed\_page} map to the kernel, indicating whether the pages associated with each pkey are sealed.
We modify the \texttt{pkey\_mprotect} system call to check the \texttt{sealed\_page} map and only allow adding new pages to a pkey domain if the associated pages of that domain are not sealed.
As shown in Figure~\ref{fig:log}, we add a new system call, \texttt{pkey\_seal(int pkey, bool seal\_domain, bool seal\_page)}, which allows the programmer to seal a domain and/or its associated pages.
Note that once a domain or its associated pages are sealed, the seal cannot be broken unless the corresponding pkey and all its associated pages are freed.

\textbf{Sealing permissions:} In this scenario, we assume \texttt{Func-D} is a third-party component, suffering from a buffer overflow vulnerability.
As shown in Figure~\ref{fig:log}, an attacker can leverage this vulnerability to inject a {\WRPKRU} instruction at run-time and modify the permission bits of the \texttt{log} to \texttt{readable-writable}.
Subsequently, the attacker can falsify the sensitive contents of the financial record.
Intel MPK does not protect pkey permissions against control-flow hijacking attacks that leverage the \texttt{WRPKRU} instruction.
To prevent such a tampering, we provide a permission sealing feature, which allows the developer to restrict the execution of the \texttt{\WRPKRU} instruction to a specified range of memory addresses.
In this example, we aim to restrict the occurrence of the {\WRPKRU} instruction to the address range of \texttt{Func-A}.

\begin{figure}[t]
  \centering
  \includegraphics[width=0.98\linewidth] {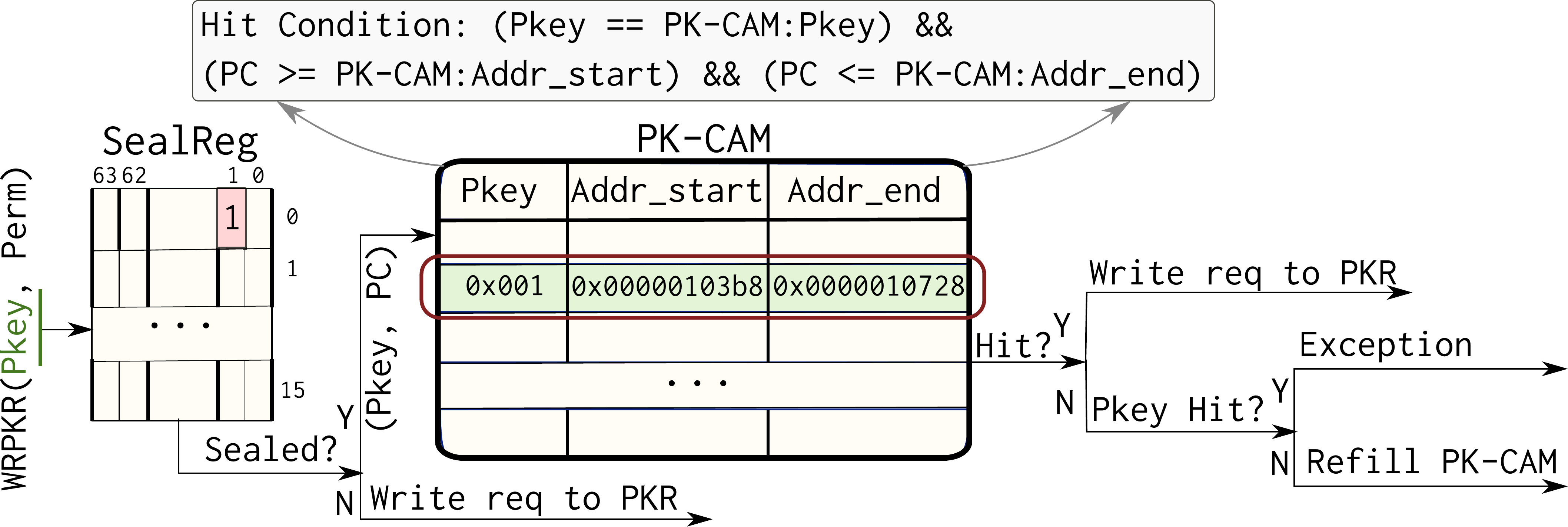}
  \caption{High-level view of {\RVMPK}'s hardware support to seal pkey permissions.}
  \label{fig:seal}
\end{figure}

\label{sec:eval}
\begin{figure*}[t]
  \centering
  \includegraphics[width=0.98\linewidth] {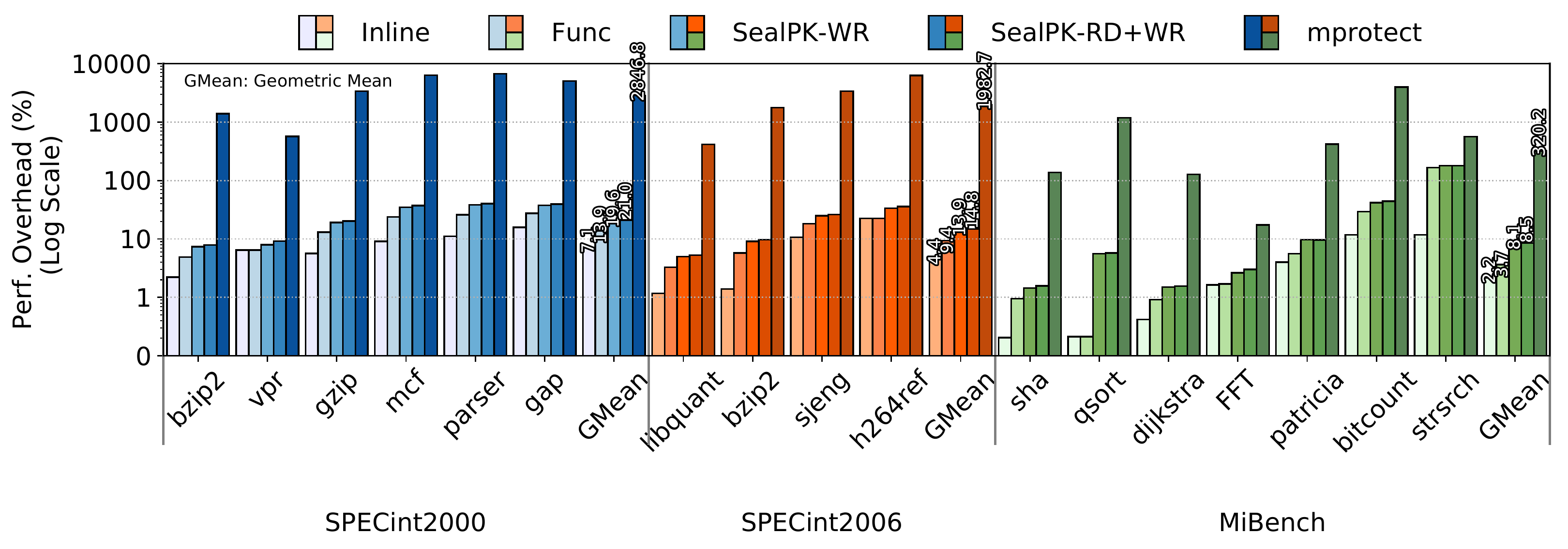}
  \caption{Performance overhead of various LLVM-based shadow stack implementations for SPECint2000, SPECint2006, and MiBench benchmarks.
  The {\Inline} implementation is a front-end LLVM pass, where the shadow stack capability is inserted as an inline code.
  The {\Func} implementation uses a function call in the front-end pass rather than an inline code.
  {\MPK} is implemented as a back-end pass built upon {\Func}, where it writes the new value of pkey permission bits without maintaining the rest of the permission bits.
  {\MPKR} adds the support to read the corresponding row of the pkey before updating it.
  {\mprotect} is implemented as an inline front-end pass by invoking {\mprotect} system call before and after writing the return address into the shadow stack.
  }
  \label{fig:performance}
\end{figure*}

At hardware level, as shown in Figure~\ref{fig:seal}, we keep track of sealed pkey permissions using a local memory, called {\SealReg}.
We modify the Rocket core's pipeline to consult {\SealReg} prior to executing a {\WRPKRU} instruction.
If the permission bits of the pkey are sealed, the {\WRPKRU} instruction is only allowed in the permissible range, specified by the developer.
We leverage a Content-Addressable Memory (CAM) like structure, named {\PKCAM}, to cache the permissible range of each pkey.
If the pkey information is available in {\PKCAM} but the current address of the {\WRPKRU} instruction is not in the permissible range, then {\RVMPK} prevents the execution of {\WRPKRU} and causes an exception.
If {\PKCAM} does not include the pkey information, we will refill {\PKCAM}.\footnote{Currently, we trigger an interrupt and insert the pkey and its permissible range to {\PKCAM} in the OS interrupt handler. 
As part of our future work, we plan to delegate this interrupt to user level and provide a secure software library to update {\PKCAM}.
}

We also provide the software support for sealing the permissions.
We provide two new custom instructions, i.e., {\sealstart} and {\sealend}, to specify the contiguous permissible range of each pkey.
Although these instructions can be added to the source code (Figure~\ref{fig:log}), the more efficient way of using them is by a compiler pass or through run-time mechanisms such as ld-preload.
After specifying the start and end addresses of a permissible range for {\WRPKRU}, the developer has to invoke a newly added system call (\texttt{pkey\_perm\_seal}) to seal the permissions.
This system call leverages a custom instruction, which is only accessible to the supervisor mode, to seal the permission bits by updating the \texttt{\SealReg} and \texttt{PK-CAM}.
We modify the Linux kernel to maintain the {\SealReg} information as well as permissible range of each pkey during context switches for each process.
Note that {\SealReg} and the permissible range of a pkey are implemented similar to a one-time fuse, i.e., they can only be written once for each process.
Hence, after configuration, the permission sealing feature cannot be modified.
The simplicity and efficiency of our permission sealing feature distinguishes our work from existing works focused on preventing the manipulation of a domain's permissions by an attacker, e.g., \cite{hedayati2019hodor} and \cite{vahldiek2019erim}.

By leveraging {\RVMPK}'s sealing features, the software developer can implement a \textbf{tamper-proof} \texttt{log} of financial records in the face of buggy and malicious third-party components.
\section{Evaluation}

\subsection{Experimental Setup}
We use the Chisel HDL~\cite{bachrach2012chisel} to implement {\RVMPK} on a RISC-V Rocket core~\cite{asanovic2016rocket}, configured with a 16KB L1 instruction and data caches.
We add the OS support for {\RVMPK} to the Linux kernel v4.15.
As a case study, we implement an isolated shadow stack using LLVM front-end and back-end passes.
We use Clang v.7 and v.8 for our front-end and back-end passes, respectively.
We prototype our hardware design with the full software stack on a Xilinx Zedboard FPGA.

For performance evaluation, we use RISC-V LLVM to cross-compile {\numSpec} applications (out of 12) from SPECint2000~\cite{henning2000spec}, {\numSpecSix} applications (out of 12) from SPECint2006~\cite{henning2006spec}, and {\numMibench} applications from MiBench~\cite{guthaus2001mibench} benchmark suites.
Due to compilation issues and memory limitations of our FPGA, we were not able to successfully cross-compile and run all the applications from these benchmark suites.
In particular, for SPECint2000, we got a segmentation fault for the baseline execution of \texttt{vortex} and \texttt{gcc}, and faced LLVM cross-compilation issues for the remaining {\numSpecRemaining} applications.
For SPECint2006, with the baseline code, we got an out of memory error for \texttt{mcf} and a segmentation fault for \texttt{gcc}.
We faced various LLVM cross-compilation issues for the remaining {\numSpecSixRemaining} applications.
Note that RISC-V LLVM is still not as mature as GCC support for RISC-V.
In our evaluations, we use the \texttt{large} inputs for MiBench and evaluate SPECint2000 and SPECint2006 applications using \texttt{test} inputs.\footnote{We use the \texttt{test} inputs for SPEC evaluations due to the memory limitation of our FPGA board (256MB) as well as the long execution time of the benchmarks for the \texttt{mprotect} comparison point (multiple days).}
\subsection{Case Study: An Isolated Shadow Stack}
To demonstrate the effectiveness of {\RVMPK}, as a case study, we use {\RVMPK} to protect an isolated shadow stack that prevents Return-Oriented Programming (ROP) attacks.
An ROP attack is a contemporary code-reuse attack that allows an attacker to execute arbitrary code by overwriting the return addresses on the stack.
A shadow stack protects the return addresses by storing them in a separate memory.
It is imperative to guarantee the integrity of the shadow stack~\cite{burow2019sok}, i.e., the shadow stack area should be an \textbf{isolated} area within the process' address space to prevent attackers from modifying it.
We isolate the shadow stack memory in a protection domain.
Once the shadow stack memory is allocated and assigned to a domain, no more pages will be added and the protection domain stays the same during the process execution.
We leverage the domain and page sealing features to protect the allocated domain and pages of the shadow stack from further modifications (similar to scenarios described in Section~\ref{sec:sealable}) after the initial configuration.

For the shadow stack implementation, we first implement a baseline front-end pass LLVM plugin~\cite{delshadtehrani2020phmon}.
This front-end pass allocates a memory area for the shadow stack and instruments the prologue and epilogue of each function to push the original return address into the shadow stack memory and pop the shadow return address from that memory, respectively.
To isolate the shadow stack, we modify the front-end pass to allocate a pkey and to assign it to the shadow stack memory pages.
To protect the shadow stack from modifications, we initialize the pkey as \texttt{read-only}.
We implement a RISC-V back-end pass to temporarily update the pkey permission to \texttt{readable-writable} in the prologue, where we push the return address into the shadow stack.
Right after pushing the return address, the back-end pass disables the pkey write permission.
Our back-end pass inserts the required {\RDPKRU} and {\WRPKRU} instructions to update the pkey's permission bits.
We can leverage our permission sealing feature to restrict the {\WRPKRU} occurrences to the memory range of the back-end pass.

In our evaluations, we measured the total execution time of an application as our performance metric.
For the baseline, we compiled the benchmarks using Clang v.8 without applying any passes and ran the benchmarks on an unmodified core and Linux kernel.
We ran each application three times and report the geometric mean of the execution times.

Figure~\ref{fig:performance} shows the performance overhead of various shadow stack implementations compared to the baseline.
{\Inline} and {\Func} are front-end LLVM passes that cannot guarantee the integrity of the shadow stack; hence, the shadow stack memory remains unprotected.
{\MPK} and {\MPKR} are isolated shadow stack implementations, leveraging {\RVMPK} in a back-end pass.
{\mprotect} is our comparison point, an isolated shadow stack implemented by leveraging the {\mprotect} system call.
As expected, using {\mprotect} incurs considerable performance overhead, i.e., {\psmprotect}, {\psixmprotect}, and {\pmmprotect}, on average, for SPEC2000, SPEC2006, and MiBench, respectively, which makes it an infeasible option.
{\mprotect} requires a context switch into the kernel, followed by a full page table walk to change the permissions of all the specified pages, and then a TLB flush.
On the contrary, leveraging {\RVMPK} to implement an isolated shadow stack uses a user-space instruction to modify the pkey permission bits.
{\MPKR}, has an average of {\psMPKR}, {\psixMPKR}, and {\pmMPKR} performance overhead for SPEC2000, SPEC2006, and MiBench applications, respectively.

\vspace{-5pt}
\subsection{Hardware Overhead of {\RVMPK}}
Table~\ref{tab:area} shows the FPGA utilization of adding {\RVMPK} to the Rocket core compared to the baseline unmodified Rocket core.
In our FPGA prototype, enhancing Rocket core with {\RVMPK} increases the LUT and FF utilization by {\LUTs} and {\FFs}, respectively.\footnote{Note that the reported LUT and FF overhead includes the resource utilization of adding RoCC custom instruction support to the Rocket core.}
The main source of area and power overhead for {\RVMPK} is {\RVPKR}, a 2Kb local memory.
Accordingly, we estimate that our power overhead is also less than 5\%, even when considering a 100\% access rate to {\RVPKR}.
In our FPGA evaluation, the Rocket core operated with a maximum
frequency of 25~MHz (both in the baseline and the enhanced version with {\RVMPK} experiments).\footnote{Note that an ASIC implementation of the Rocket core can perform with a target frequency of 1~GHz.}
According to our FPGA place and route results, {\RVMPK}'s modifications to the Rocket core did not change the critical path.

\begin{table}[!t]
\caption{The FPGA utilization of {\RVMPK} compared to the baseline Rocket core.}
\resizebox{0.98\linewidth}{!}{
\begin{tabular}{l|c|c|c|c|}
\cline{2-5}
 & \multicolumn{2}{c|}{Baseline}
 & \multicolumn{2}{c|}{Rocket Core + {\RVMPK}} \\ \cline{2-5}
 & \multicolumn{1}{l|}{Used}
 & \multicolumn{1}{l|}{Utilization}
 & \multicolumn{1}{l|}{Used} & \multicolumn{1}{l|}{Utilization} \\ \hline
\multicolumn{1}{|l|}{Total Slice Luts} & 32030 & 60.21& 35019 & 65.83 \\ \hline
\multicolumn{1}{|l|}{Luts as logic} & 30907 & 58.1 & 33852 & 63.63  \\ \hline
\multicolumn{1}{|l|}{Luts as Memory}  & 1123  & 6.45  & 1167 & 6.71 \\ \hline
\multicolumn{1}{|l|}{Slice Registers as Flip Flop} & 16506 & 15.51 & 19392 &	18.23\\ \hline
\end{tabular}}
\label{tab:area} \vspace{-5pt}
\end{table}
\section{Related Work}
\label{sec:relatedwork}
There is a considerable amount of prior work on intra-process memory isolation.
A Software Fault Isolation (SFI) technique~\cite{wahbe1993efficient} instruments each memory access by address masking instructions to prevent unintended memory accesses; however, it suffers from large performance overhead.
Prior to Intel MPK, CODOMs~\cite{vilanova2014codoms} and CHERI~\cite{watson2015cheri} proposed efficient capability-based systems, which require significant and invasive hardware modifications.
IMIX~\cite{frassetto2018imix} enables secure data encapsulation by minimally extending the x86 ISA with secure load and store instructions.
To address Intel MPK's limitations, Hodor~\cite{hedayati2019hodor} and ERIM~\cite{vahldiek2019erim} combine Intel MPK with binary inspection to prevent reusing of \texttt{WRPKRU} instruction by an attacker.
The sealing permission feature of {\RVMPK} provides a similar capability by restricting valid {\WRPKRU} instructions to a contiguous range of memory addresses for each pkey.
Although our sealing feature is limited to one valid memory range for each pkey, its simplicity and efficiency distinguishes our work form Hodor and ERIM.
To allow the occurrence of {\WRPKRU} instructions in more than one trusted component, we can  rely on a CFI technique for the RISC-V Rocket core~\cite{canakci2020efficient} to protect {\RVPKR} from manipulation by an attacker.
libmpk~\cite{park2019libmpk} and Xu et al.~\cite{xu2020hardware} provide a software-based and a hardware-based virtualization technique, respectively, to address the limited number of pkeys.
We can leverage such virtualization techniques to support more than 1024 domains for {\RVMPK}.

Donky~\cite{schrammel2020donky} provides a secure user-space software framework to protect the domain permissions against CFI attacks without relying on binary inspection or CFI.
Donky proposes a pkey extension for RISC-V ISA, and implements it on the Ariane core~\cite{ariane2020}.
Similar to {\RVMPK}, Donky uses the 10 unused bits of Sv39 PTEs to store the pkeys; however, Donky relies on a 64-bit CSR (managed by a software library) to store the permission bits of only 4 pkeys at a time.
If the pkey of the accessed memory address is not loaded into that CSR, Donky requires extra cycles for the software library (which stores all the pkey information) to load the missing pkey and its permission into the register.
In our design, we access {\RVPKR} in the same cycle as page-table permission checks.
The permission sealing feature of {\RVMPK} allows us to protect a domain against CFI attacks in cases where the valid {\WRPKRU} instructions occur in contiguous memory addresses.
In addition to this feature, {\RVMPK} provides two other novel sealing features to prevent a domain and its associated pages from tampering.
\vspace{-4pt}
\section{Conclusion}
In this paper, we proposed an efficient intra-process memory isolation technique ({\RVMPK}) for a RISC-V processor, which supports up to 1024 domains.
In our design, we provided three novel sealing features to protect a domain, its associated pages, and its permission bits from unauthorized modifications.
To address the pkey use-after-free problem, we used an OS-level lazy de-allocation approach.
We prototyped RISC-V Rocket + \RVMPK~on an FPGA with full software stack, and demonstrated the efficiency of {\RVMPK} by securing a shadow stack.
\vspace{-5pt}
\section{Acknowledgments}
This material is based upon work supported by the National Science Foundation under Grant No. CNS-1916393.

\bibliographystyle{plain}
\bibliography{sealpk-2020}
\end{document}